\documentclass[9pt,Preprint]{lapreprint}

\usepackage[version=4]{mhchem} 
\usepackage{siunitx}    
\usepackage{pdflscape}  
\usepackage{rotating}   
\usepackage{textgreek}  
\usepackage{gensymb}    
\usepackage[misc]{ifsym} 
\usepackage{orcidlink}  
\usepackage{listings}   
\usepackage{colortbl}   
\usepackage{tabularx}   
\usepackage{longtable}  
\usepackage{subcaption}
\usepackage{multirow}
\usepackage{snotez}     
\usepackage{csquotes}   

\DeclareSIUnit\Molar{M}

\usepackage[			
	backend=biber,      
    style=authoryear,   
	natbib=true,		
	hyperref=true,	    
	alldates=year,      
]{biblatex}

\AtEveryBibitem{
    \clearfield{urlyear}
    \clearfield{urlmonth}
    \clearlist{language}
}

\title{A Causal Framework for Precision Rehabilitation}

\author[ \orcidlink{0000-0001-5714-1400} 1, 2 \Letter]{R. James Cotton}
\author[3, 4]{Bryant A. Seamon}
\author[5, 6, 7]{Richard L. Segal}
\author[5, 6]{Randal D. Davis}
\author[8]{Amrita Sahu}
\author[9]{Michelle M. McLeod}
\author[1, 2]{Pablo Celnik}
\author[10, 11]{Sharon L. Ramey}

\affil[1]{Shirley Ryan AbilityLab}
\affil[2]{Northwestern University, Department of Physical Medicine and Rehabilitation}
\affil[3]{Division of Physical Therapy, Department of Rehabilitation Sciences, College of Health Professions, Medical University of South Carolina}
\affil[4]{Ralph H. Johnson VA Healthcare System, Charleston, SC}
\affil[5]{Department of Health Sciences and Research, College of Health Professions, Medical University of South Carolina}
\affil[6]{National Center of Neuromodulation for Rehabilitation, Medical University of South Carolina}
\affil[7]{NIH/NICHD/NCMRR Medical Rehabilitation Research Resource Network National Coordinating Center}
\affil[8]{Department of Physical Medicine and Rehabilitation, School of Medicine, University of Pittsburgh}
\affil[9]{Arthritis Foundation}
\affil[10]{Fralin Biomedical Research Institute, Virginia Tech}
\affil[11]{VTC School of Medicine, Virginia Tech}

\metadata[]{\Letter\hspace{.5ex} For correspondence}{rcotton@sralab.org}
\metadata[]{Keywords}{rehabilitation; precision rehabilitation; causal inference; international classification of functioning; rehabilitation treatment specification system; computational neurorehabilitation}

\leadauthor{R. James Cotton}
\shorttitle{Causal Precision Rehab}

\begin{document}
\maketitle

\begin{abstract}
Precision rehabilitation offers the promise of an evidence-based approach for optimizing individual rehabilitation to improve long-term functional outcomes. Emerging techniques, including those driven by artificial intelligence, are rapidly expanding our ability to quantify the different domains of function during rehabilitation, other encounters with healthcare, and in the community. While this seems poised to usher rehabilitation into the era of big data and should be a powerful driver of precision rehabilitation, our field lacks a coherent framework to utilize these data and deliver on this promise. We propose a framework that builds upon multiple existing pillars to fill this gap. Our framework aims to identify the Optimal Dynamic Treatment Regimens (ODTR), or the decision-making strategy that takes in the range of available measurements and biomarkers to identify interventions likely to maximize long-term function. This is achieved by designing and fitting causal models, which extend the Computational Neurorehabilitation framework using tools from causal inference. These causal models can learn from heterogeneous data from different silos, which must include detailed documentation of interventions, such as using the Rehabilitation Treatment Specification System (RTSS). The models then serve as digital twins of patient recovery trajectories, which can be used to learn the (ODTR). Our causal modeling framework also emphasizes quantitatively linking changes across levels of the International Classification of Functioning, Disability, and Health (ICF) to ensure that interventions can be precisely selected based on careful measurement of impairments while also being selected to maximize outcomes that are meaningful to patients and stakeholders. We believe this approach can provide a unifying framework to leverage growing big rehabilitation data and AI-powered measurements to produce precision rehabilitation treatments that can improve clinical outcomes.
\end{abstract}

\section{Introduction}

Precision Rehabilitation is an emerging paradigm that promises an evidence base to prescribe optimized, personalized treatments to improve function and independence. Precision Rehabilitation aims to mirror the mission championed by Precision Medicine; ``to provide the right treatment to the right person at the right time'' \parencite{denny_precision_2021}. However, there are distinct differences between medical and rehabilitation care, necessitating a broader vision for Precision Rehabilitation. Rehabilitation is typically focused on patient-centered goals, with the desired outcome being an improvement in function and independence over the patient's lifespan. The World Health Organization's International Classification of Functioning, Disability, and Health (ICF) \parencite{world_health_organization_international_2001} frames the complex interplay among physiology, personal characteristics, and social systems with function and participation. This complexity increases when one considers the important need to understand an individual's lifespan trajectory and current place in their rehabilitation continuum.

Advancing Precision Rehabilitation requires strategically stimulating research and clinical efforts to address the complexity of rehabilitation while considering the comprehensive needs of an individual. Rehabilitation is well positioned to take this step, given the standard practice of interprofessional collaboration to navigate the complexity surrounding function. However, specific efforts are still needed to further our understanding of and ability to address individual differences concerning various functional outcomes, the treatment and assessment process, and patient-reported engagement toward the development and achievement of goals. This is not a trivial endeavor and will require a well-defined and robust framework to guide researchers and clinicians to create and deliver more effective interventions with enduring benefits. \textcite{french_precision_2022} outlined some necessary components for precision rehabilitation, including using synergistic study designs, standardization of measurement, implementation of precise measures for function across the lifespan, development of comprehensive databases, and formation of interprofessional teams.

These components are necessary but not sufficient. To advance Precision Rehabilitation, \textit{it is also essential to establish clear and explicit linkages between interventions, the plasticity they induce and physiologic mechanisms they act on, and the changes in function they facilitate from the level of impairment to participation}. This linkage will allow stakeholders, clinicians, and researchers to establish a scientifically-based framework that links changes in impairment with changes in functional outcomes in a way that informs treatment prescriptions and study designs and determines the ``right treatment for the right person at the right time.'' More formally, what we desire for Precision Rehabilitation is the Optimal Dynamic Treatment Regimens (ODTR) \parencite{murphy_optimal_2003}. This evidence-based model, based on the data available about a patient, identifies the subsequent intervention to maximize attaining their goals or the next test required to determine this. Determining the ODTR can be accomplished by building upon the framework of Computational Neurorehabilitation \parencite{reinkensmeyer_computational_2016}, highlighting the importance of computationally modeling plasticity rules. We extend this to emphasize modeling the influence of rehabilitation interventions across the levels of the ICF and over time. This makes it explicit that our goal is to optimize long-term function through a better understanding of impairments, recovery mechanisms, and important factors from social determinants of health to genetics. The integration of different measurements to support decision-making should be well grounded in measurement theory and construct validity, and measurements should be rigorously linked and validated for the different classes of biomarkers established by the NIH and FDA \parencite{fda-nih_biomarker_working_group_best_2016}. Our framework also embraces the Rehabilitation Treatment Specification System (RTSS) \parencite{hart_theory-driven_2019, van_stan_rehabilitation_2019}, where clinicians document specific targets of interventions and the active ingredients of interventions that are thought to be therapeutic. The RTSS also recommends documenting the hypothesized mechanisms through which active ingredients improve their treatment targets. This aligns well with our framework, which recommends all this data be cohesively modeled using the tools of Causal Inference (CI) \parencite{pearl_causality_2009, pearl_book_2018}. This will allow us to incorporate the known mechanistic recovery processes while spanning this broader perspective powered by big data from rehabilitation. These causal models can then serve as Digital Twins \parencite{bjornsson_digital_2019, kamel_boulos_digital_2021, masison_modular_2021} that are used to identify the ODTR.

Our framework is designed pragmatically to advance precision rehabilitation using the big data that is becoming available in rehabilitation. In addition to traditional clinical outcome assessments, this includes precisely measuring movement, both during rehabilitation and in the community, produced from AI-powered computer vision analysis and wearable sensor data. Causal modeling allows learning from heterogeneous datasets from multiple sites, which may include different patient populations, treatment patterns, and interventions. The tools of CI can leverage this heterogeneity to identify the most effective interventions. Computational approaches like Federated Learning \parencite{li_federated_2020, nguyen_novel_2022} can also help to learn these models from data at multiple sites without requiring individual sites to share their data through centralized repositories, reducing many privacy concerns and barriers to collaboration.  \textit{This big-data approach can mitigate the infeasibility of performing all the randomized controlled trials (RCTs) necessary to optimize our interventions, particularly combinatorial interventions.}

In this paper, to make the concepts more specific, we primarily focus on motor rehabilitation after a neurologic injury. However, our framework is naturally extensible to all aspects of rehabilitation and patient populations. Additionally, describing this framework is a first step. Developing the causal models that we believe are necessary is a significant research endeavor.

The need to adopt a unifying transdisciplinary framework to guide research and practice was a key theme that emerged from the National Institutes of Health (NIH)-supported consortium of six Medical Rehabilitation Research Resource centers' MR3 Network Scientific Retreat on Precision Rehabilitation. The goal of the Retreat was to delineate a future course of scientific inquiry and knowledge synthesis, building upon successes in other areas of human health and identifying emerging discoveries in medical rehabilitation. Other key themes that emerged from the Retreat were 1) the high value of patients as active collaborators in research and treatment, 2) the high impact of incorporating biobehavioral markers as well as genomics, and 3) the need for a unifying computational framework \parencite{cotton_letter_2022}.
The purpose of this paper is to present a Causal Framework for Precision Rehabilitation. Throughout this paper, we incorporate the integral components outlined in \textcite{french_precision_2022}, subsequent letters \parencite{cotton_letter_2022, french_response_2022}, prior frameworks and consensus statements  \parencite{reinkensmeyer_computational_2016, hart_theory-driven_2019, van_stan_rehabilitation_2019, kwakkel_standardized_2019, van_criekinge_standardized_2023}, insights that emerged from the MR3 Network Precision Rehabilitation Retreat, and our work developing an outline for this framework. We expect that this framework will enable the delivery of an integrated approach that will guide research activities and changes in clinical care that are consistent with a precision rehabilitation approach.

\section{Precision Rehabilitation: Concept and Necessity}

Precision Rehabilitation requires a broader conceptual approach than Precision Medicine to allow researchers, clinicians, and policymakers to produce the necessary advances to identify and deliver the ``right treatment, to the right person, at the right time.'' This framework must describe how an individual's function changes over time. Function is commonly organized under the ICF \parencite{world_health_organization_international_2001} specifically, in three domains: 1) impairments in body structure and function, such as weakness or spasticity; 2) activity limitations, such as walking and balance performance; and 3) participation restrictions, such as the inability to safely ambulate in the community and visit family. While performance in each domain is influenced by the person's health condition (the medical model), they are also influenced by social and environmental factors (the biopsychosocial model). Current precision medicine frameworks do not attempt to model these interactions, and computationally formalizing these connections will benefit the rehabilitation community greatly. Many rehabilitation researchers focused more on impairments than participation because they are relatively easier to measure, more closely linked to mechanisms researchers study, and often respond more sensitively to interventions. However, clinicians, patients, family, and friends are typically more concerned with participation and real-world functioning. Knowing whether an intervention will reduce impairment sufficiently to change function in a way that matters to a patient is critical when deciding if it should be deployed into clinical practice versus needing to be optimized further as a research endeavor.

Rehabilitation has historically struggled to precisely measure many of our treatment targets, from movement quality to performance in the real world. For example, given how frequently we treat walking, it is shocking that we primarily use only a stopwatch to measure it. \textit{In lieu} of direct measurements, the field has created a veritable forest of outcome measures that seek to capture different aspects of function. However, these do not always generalize as intended to real-world function \parencite{major_validity_2013, doman_changes_2016}. The reliance on classical test theory for patient-reported and clinician-observed outcomes development caused sample characteristics, rather than the underlying constructs or phenomenon of interest, to be the focus of measurement. This has made it difficult to meaningfully test causal relationships of underlying constructs as potential treatment mechanisms. Our framework will outline how innovative approaches to measurement can help elucidate the underlying constructs for activities and participation, which can be used to design interventions and inform clinical trials. This will be augmented by recent advances in wearable sensors and computer vision, and are now expanding access to detailed movement data and the ability to monitor real-world behavior \parencite{patel_review_2012, kristoffersson_systematic_2022, stenum_clinical_2023, bandini_measuring_2022, cimorelli_validation_2024}. In many cases, these data are used to automate existing outcome measures, which can reduce labor and increase reliability. However, automating assessments to obtain more of the data we have been collecting for years is unlikely, by itself, to produce fundamental new insights into rehabilitation or improve patient-valued outcomes. Our proposed framework is designed to determine which features from these new data sources give us greater insight into activities and participation and facilitate identifying patient subgroups needing different treatments at distinct times along the rehabilitation continuum. Furthermore, it provides a cohesive framework to test the construct validity of different measurements or their combinations and relate them to underlying causal mechanisms.

Another barrier to precision rehabilitation is that it is impossible and inefficient to perform randomized controlled trials to explore all the possible combinatorial interventions and patient characteristics to understand this complex process. We believe that methods to leverage Big Data will be necessary to address these questions. However, this will require a framework that quantitatively models functioning across levels of the ICF from impairments to participation, how these change over time, and the influence of different interventions. This framework should also be able to build upon and benefit from the work studying plasticity throughout the nervous system, extending the framework of Computational Neurorehabilitation \parencite{reinkensmeyer_computational_2016} to a broader framework for Precision Rehabilitation. Below, we will elaborate on how the field of Causal Inference can enable this framework and allow us to learn from large, longitudinal datasets from rehabilitation. This will also require the longitudinal data to be accompanied by detailed documentation of treatment interventions. This is frequently lacking in both clinical documentation and even research studies. Treatment components documented according to the Rehabilitation Treatment Specification System (RTSS) \parencite{hart_theory-driven_2019, van_stan_rehabilitation_2019} should include three elements: the treatment target, the active ingredients thought to induce change in the target, and the hypothesized mechanism(s) of these interventions. Thus, both Computational Neurorehabilitation and the RTSS share our focus on causal mechanisms. In our framework, Causal Inference (CI) \parencite{pearl_causality_2009, pearl_book_2018} is the computational engine that can help support or reject these mechanistic hypotheses, as well as provide a tool to generate new hypotheses.

\section{Key Components of Precision Rehabilitation}

The two pillars of our proposed framework are the causal models and the data on which these will be trained. We elaborate more on causal models and inference below, but at a high level, it is the combination of formalizing causal hypotheses of rehabilitation and recovery with causal models and then fitting them to data, which can overcome the nihilism of ``correlation is not causation.'' Following approaches in machine learning on training Foundation Models \parencite{bommasani_opportunities_2021}, we envision properly structured causal models will continue improving as the amount of data they are trained on increases, showing an ability to absorb knowledge from multiple sources of heterogeneous data. \textcite{french_precision_2022} highlight the importance of including standardized functional measurements and sharing data between different silos. One standardization encouraged through the NIH data sharing requirements is the NIH Common Data Elements (CDE) \parencite{grinnon_national_2012}. However, these codes are frequently inadequate. For example, when characterizing gait, there is a CDE for daily step count and fall rate, but there are none for spatiotemporal gait parameters or other quantitative gait measures. ICF codes can provide a standardized way to refer to some functions, such as the body function domain of ``b770 Gait pattern function'' and activities and participation domain of ``d4500 Walking short distances.'' However, these are also fairly broad, and the codes do not specify how each construct is measured. Furthermore, performance on any outcome measure could be influenced by performance across multiple domains. Thus, there is a dual problem of standardizing over a rapidly expanding set of measurements and disentangling the causal structure within the ICF constructs.  Below, we discuss other data that must be included and expand this to incorporate concepts from biomarkers \parencite{fda-nih_biomarker_working_group_best_2016}.

Rich rehabilitation datasets will be deeply interwoven with protected health information (PHI), introducing huge barriers to large-scale data sharing. One approach around this may be sharing deidentified slices of data or aggregated statistics, but this will limit learning the detailed causal structure of the data. Federated learning \parencite{li_federated_2020, nguyen_novel_2022} is a promising alternative, where models are learned in a decentralized manner without data ever leaving any given site. Implementation of this places demands on each site to have the appropriate secure computational infrastructure to perform this learning. The causal discovery methods discussed in the Details~on~Causal~Inference can also be performed using Federated Learning, which are approaches that train a model on multiple datasets without requiring those datasets to be centrally shared \parencite{abyaneh_fed-cd_2023}. Standardizing both the data formats and the Federated Learning architecture for deployment across different electronic medical records will be a substantial endeavor but will yield large dividends for the future of precision rehabilitation. Another promising alternative is using AI to generate synthetic datasets that capture the clinically pertinent aspects of the data while removing any identifiable data \parencite{kweon_publicly_2023}. While this has shown promise for some narrow tasks, scaling this to the complexity of rehabilitation data essentially presumes first learning the causal model we are describing.

\subsection{Optimal Dynamic Treatment Regimen (ODTR)}\label{odtr}

\begin{figure}[!htbp]
\centering
\includegraphics[width=1\linewidth]{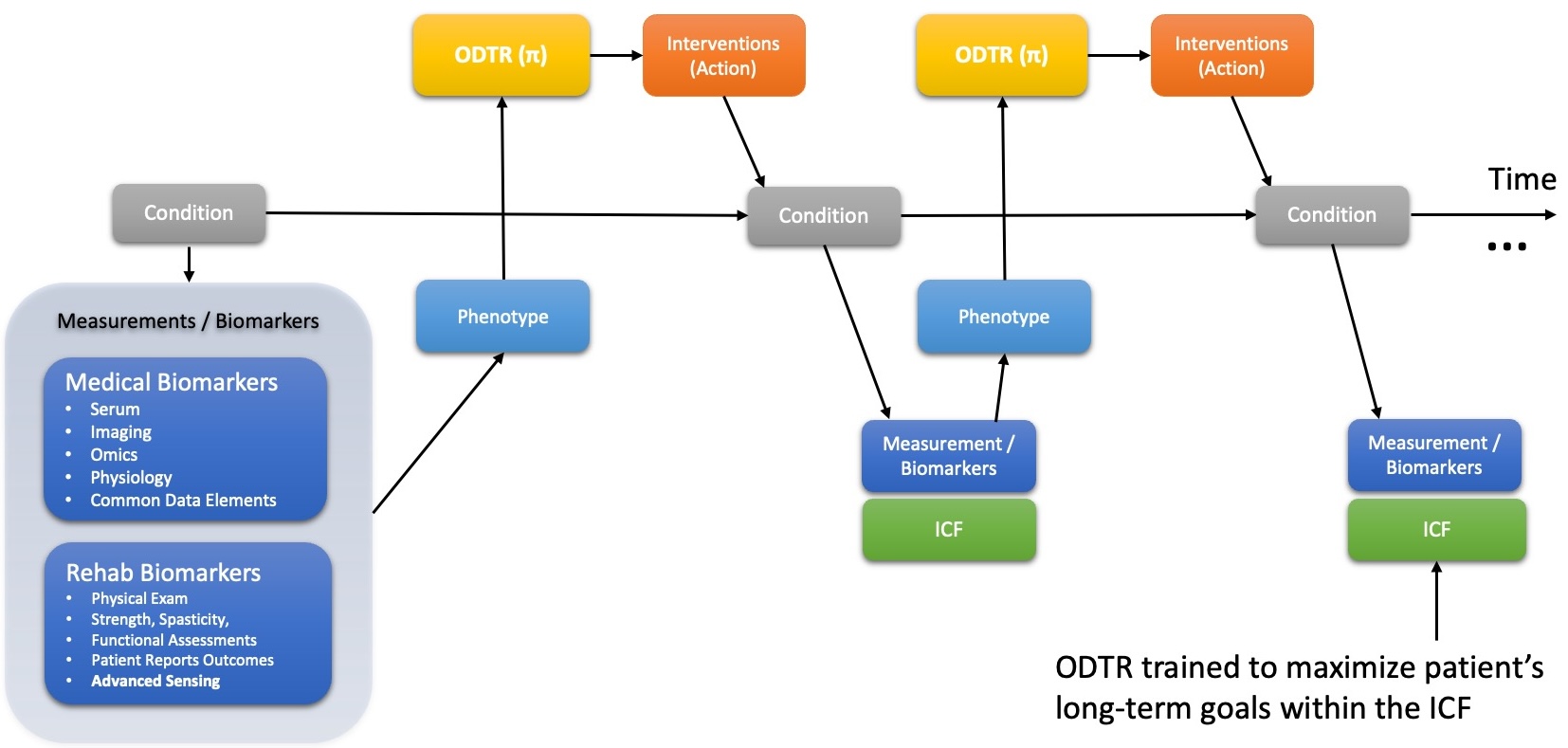}
\caption[]{High-level diagram of how the Optimal Dynamic Treatment Regimen (ODTR) would be applied longitudinally (iteratively) to determine the optimal intervention at any stage during rehabilitation. At each point in time, the phenotype of a patient would be established based on the set of measurements and biomarkers acquired to date. This phenotype would be passed to the decision policy, predicting the next intervention(s) most likely to maximize long-term function. This would be applied, and the process would be repeated. Critically, the ODTR is designed not only to make some short-term change in impairment but also to maximize function and participation over the patient's lifespan, aligned with the (potentially changing) goals of the patient.}
\label{fig:odtr}
\end{figure}

The ultimate goal of precision rehabilitation is a prescription for the optimal interventions to maximize highly-valued, long-term outcomes. However, the space of treatment strategies is too large to be feasibly tested through traditional clinical trials. This makes identifying the optimal interventions for a given individual very challenging. Furthermore, rehabilitation is an evolving process, not a singular thing administered once. Decisions must be made repeatedly about interventions, such as determining the active therapy ingredients to include today. This sequential decision-making is often referred to as a dynamic treatment regimen, and the goal of precision rehabilitation can be formalized as identifying the Optimal Dynamic Treatment Regimens (ODTR) \parencite{murphy_optimal_2003}. Despite the close match to the process of rehabilitation, an ODTR formalism has rarely been used in rehabilitation. A recent exception was optimizing decision strategies for managing lower back pain \parencite{naumzik_data-driven_2023}.

Figure~\ref{fig:odtr} shows what using a learned ODTR would look like. At any point in time, a set of measurements are available, which are used to phenotype the patient. This phenotype is then passed to the ODTR, which suggests the subsequent intervention(s) predicted to maximize long-term function. This process is then iterated after obtaining the next set of measurements. This could be performed at a range of time scales, even down to optimizing per-repetition task difficulty to balance effectiveness with the motivation and engagement of the participant.

\subsection{Theory-driven measurement}\label{measurement_theory}

Precision rehabilitation will require precise measurement. Accomplishing this goal will require a paradigm shift in measurement research to better establish construct validity and common metrics for advancing science related to activities and participation under the ICF model \parencite{pendrill_quality_2019}. Traditional measurement research or psychometrics has relied on Classical test theory for developing patient-reported or clinician-observed outcome measures. Classical test theory relies on aggregating a sample's responses to a group of items believed to describe a construct or phenomenon of interest and correlates it to a `gold standard' \parencite{cappelleri_overview_2014}. For measures of activity or participation, the `gold standard' is often other metrics thought to quantify the same desired construct. For example, correlations between one sample's score on a novel balance test with scores on the Berg Balance Scale establish construct validity for measuring balance. Unfortunately, this yields little insight into balance as the actual phenomenon of interest \parencite{mckenna_measurement_2019}, where the phenomena of interest are more commonly the likelihood of falls in the real world. One is left knowing that one samples' scores were associated with another on a different test. Only once we understand the phenomenon being measured can we begin to reach true construct validity and identify the underlying constructs as mechanisms for treatment targets \parencite{stenner_testing_1982, melin_construct_2021}.

Rasch measurement theory provides a way to study the phenomenon of interest by quantifying it from observed human behavior. The Rasch model compares an individual's responses to a set of items with a probabilistic model, where parameters represent theoretical properties of measurement \parencite{bond_applying_2015, andrich_understanding_2002}. Thus, the model can elucidate a hierarchy of item difficulty with respect to a person's ability on the same linear scale of the desired phenomenon \parencite{pendrill_quality_2019, melin_construct_2021}. Arguably, this hierarchy of the items represents a more ideal `gold standard' because it quantifies the observable phenomenon of interest (i.e., item difficulty measures). Therefore, studying the item hierarchy rather than the sample can then be used to evaluate the construct validity of the desired phenomenon. The methodology for this is to develop mathematical models that quantify the associations between item characteristics or theorized mechanisms as underlying constructs of the phenomenon with the item difficulty measures.

The most well-known example of this is seminal work by \textcite{stenner_testing_1982, stenner_measuring_2023} to develop the Lexile for measuring reading difficulty and ability. Stenner and colleagues found that mathematically representing syntax and semantics can explain 85\% of the variance in reading passage difficulty \parencite{stenner_measuring_2023}. Our group has demonstrated that this methodology can explain 87\% of the variance in gross motor task difficulty during development by mathematically representing movement requirements, body position, and the amount of support required for a task \parencite{seamon_defining_2024}. Not only does this approach give an equation from which to test relationships between underlying constructs and a phenomenon, but it also becomes the means for measuring analogous to methods used in the physical sciences \parencite{seamon_defining_2024, stenner_measuring_2023, melin_construct_2021}. Furthermore, the equation can be anchored to common reference points for developing a measurement scale with a standardized, reproducible unit \parencite{stenner_measuring_2023}.

Common metrics and equations for measurement will provide the foundation by which to build causal models that allow us to understand the relationship between activity and participation and the mechanisms that influence them. Additionally, with equations for measurement, the ability to measure more precisely can be accomplished by testing additional underlying constructs or quantifying the underlying constructs more precisely \parencite{stenner_measuring_2023}. Given the advances in novel measurement tools for more detailed quantification of movement and biological variables, this is likely to be realized in rehabilitation \parencite{stenner_measuring_2023}.

\subsection{Novel approaches for quantifying movement}\label{novel_measurement}

A driver of precision medicine has been the detailed measurement of the organ system being treated, where labs and imaging provide detailed snapshots of function to guide interventions. Similarly, in oncology, advances in genetic sequencing have been essential to advancing precision cancer treatment targeting individual cancer variations. This has been a stumbling block for rehabilitation, as we lack easy-to-use, sensitive tools to measure responses to our interventions. Particularly for motor rehabilitation, such as treating gait impairments, we more frequently measure with a stopwatch than quantitatively characterizing gait kinematics and walking quality. This is analogous to a cardiologist managing heart function with only a pulse and losing access to electrocardiograms and cardiac imaging to measure functioning!

Precise measurements are essential for precise treatment. They will allow for more precise treatment prescriptions and enable us to identify patient subgroups that respond differently to a particular intervention. They can also provide a sensitive response signal to empower research. Consensus guidelines in stroke have advocated for the need for routine kinematic measurements in the upper extremity and more recently for gait \parencite{kwakkel_standardized_2019, van_criekinge_standardized_2023} while also noting the challenges obtaining these measurements in routine practice. While new technologies are reducing these measurement barriers, it is unclear how to use this data to improve clinically meaningful outcomes. This is a gap our framework seeks to address.

In ``Illuminating the dark spaces of healthcare with ambient intelligence,'' \textcite{haque_illuminating_2020} outlines a vision where technology throughout a healthcare system captures many previously unmeasured variables. In rehabilitation, we must illuminate the dark spaces of function in the real world. Advances in wearable sensors and AI-powered computer vision now greatly reduce these barriers to quantitatively characterizing movement. Reviewing this work is beyond the scope of our article, so we refer readers to some reviews: \parencite{patel_review_2012, kristoffersson_systematic_2022, stenum_clinical_2023}. To date, most work with these technologies has focused on algorithm development and automating existing outcome measures (e.g., \textcite{adans-dester_enabling_2020}). An important question that our precision rehabilitation framework seeks to determine is the theory underlying our measurements, what must be measured, and what accuracy is required to model recovery and optimize care.

Novel measurement devices have varying accuracies and complementary strengths. For example, a single or sparse set of wearable sensors can only track a limited number of body segments or collect important statistics like step count and timing. Our team has developed a plurality of tools and algorithms to make it easier to characterize movement in the clinic and community, such as reconstructing kinematics from multiple cameras or a smartphone camera optionally combined with an arbitrary set of wearable sensors \parencite{cotton_improved_2023, cotton_optimizing_2023, cotton_differentiable_2024, peiffer_fusing_2024, cimorelli_validation_2024} (shown later in Figure~\ref{fig:stroke_gait}C). These approaches now even allow precisely tracking every joint in the hand, which is very challenging given the high degrees of freedom of the arm and hand \parencite{firouzabadi_biomechanical_2024}. However, obtaining this data is only the start. The clinically relevant challenge is using it to improve patient outcomes. This requires determining what tasks should be measured, how accurately they must be measured, and the most practical way to obtain this data. Furthermore, it requires us to know how to choose interventions using these measurements via the ODTR. By fitting unified causal models with this data, long-term outcome data, and clinical outcome assessments, our framework can help to answer these questions.

\subsection{Rehabilitation Biomarkers}\label{biomarkers}

A key component of a precision rehabilitation framework is the identification of patient subgroups or spectrums with different prognoses or those who would benefit from different interventions. Biomarkers are essential for this goal.  They are defined as a ``characteristic that is measured as an indicator of normal biological processes, pathogenic processes or responses to an exposure or intervention'' \parencite{fda-nih_biomarker_working_group_best_2016}. Several classes of biomarkers identified by an FDA-NIH task force \parencite{fda-nih_biomarker_working_group_best_2016} are important to consider in our framework, including 1) diagnostic biomarkers, which indicate the presence of disease or disease subtype; 2) response biomarkers, which measure responses to interventions; 3) prognostic biomarkers, which predict the future course of a disease; and 4) predictive biomarkers, the presence of which predict a different response to a specific intervention. Biomarkers can belong to multiple classes but must be validated for each context of use. To fully leverage biomarkers in rehabilitation, it is important to cast a wide tent that includes biobehavioral, social, and genetic factors. While we primarily focus on motor rehabilitation in this manuscript, it will be important to include biomarkers that capture cognition, communication, and mental health.

As an example suggestive of a movement-based biomarker, in \textcite{cotton_self-supervised_2023}, we found that self-supervised learning applied to a large dataset of gait samples learned an embedding that can classify people with a history of stroke (diagnostic biomarker) and that detected changes in gait during inpatient rehabilitation (response biomarker). Similarly, the General Movement Assessment (GMA) \parencite{einspieler_qualitative_1997} is a movement-based prognostic biomarker developed for infants to identify the risk of developing cerebral palsy. In some cases, rehabilitation biomarkers may be causally related to an outcome of interest and a modifiable treatment target. For example, \textcite{grabiner_developing_2021} found that trunk angular velocity after a stumble predicts the likelihood of falling, mechanistically contributes to causing these falls, and can be modified with training. In this case, trunk angular velocity is a predictive biomarker for a positive response to perturbation interventions. It also serves as a response biomarker to the efficacy of that intervention. In cases like this, where biomarkers are causally linked to meaningful endpoints (e.g., falls), they may also be able to serve as surrogate endpoints for clinical trials \parencite{fda-nih_biomarker_working_group_best_2016}.

Movement-based biomarkers are now being accepted as surrogate endpoints in clinical trials. For example, the 95\% stride velocity (SV95C) measured with wearable sensors in the community is a secondary endpoint in Duchenne's muscular dystrophy trial \parencite{servais_stride_2022}. In this case, studies established that SV95C is correlated with the 6-minute walk test while being more sensitive to interventions. It will be critical to continue to rigorously validate and relate these novel sensor-based biomarkers, and to contextualize them within the ICF. For example, studies have found a dissociation between changes in the Action Research Arm Test (ARAT) and upper extremity arm use at home measured with wrist-worn accelerometry \parencite{doman_changes_2016}. However, it is unclear if the metric used, distribution of accelerations, is sensitive to functional changes that might matter to a patient, such as gaining the ability to successfully perform even a limited number of bimanual tasks during the day. This highlights the importance of obtaining high-quality ground truth data of function in the real world to validate methods that can more practically be used, like wearable sensors.

Per \textcite{fda-nih_biomarker_working_group_best_2016}, biomarkers do not assess how a person feels, functions, or survives. These outcomes are instead measured with clinical outcomes assessments (COAs), including Patient-reported outcome (PRO) measures, Observer-reported outcome (ObsRO) measures, Clinician-reported outcome (ClinRO) measures, and Performance~outcome (PerfO) measures. In rehabilitation, COAs are also typically classified with the ICF domains. We anticipate for rehabilitation, some COAs or other quantitative measures of functioning may have psychometric properties that fit some of the classes of biomarkers, which may create some tension in the nomenclature of the working group. For example, speed on the 10-meter walk test (activity domain, ``d4500 Walking short distances'') is a commonly used response biomarker to monitor progress with therapy. Walking speed can also serve as a prognostic biomarker, where a walking speed after stroke between 0.4 and 0.8 meters per second predicts limited community mobility (participation domain), with speeds below 0.4 m/s predicting limited household ambulation \parencite{perry_classification_1995, bowden_validation_2008}. However, in general, there is limited work formalizing rehabilitation outcomes as biomarkers or quantitatively linking them across ICF domains and time, which can be achieved through our proposed framework.

\subsection{Rehabilitation Treatment Specification System (RTSS)}\label{rtss}

Accurate and detailed documentation of therapeutic intervention to accompany longitudinal rehabilitation outcomes data is essential to our causal framework. This type of documentation has historically been lacking, to the extent that systematic reviews by Cochrane are hindered by the lack of clarity around study interventions \parencite{french_repetitive_2016}. Documentation in clinical practice struggles similarly. The Rehabilitation Treatment Specification System (RTSS) \parencite{hart_theory-driven_2019, van_stan_rehabilitation_2019} is a framework to improve this and aligns with the goals of our proposed framework. The RTSS recommends documenting three aspects of each treatment component: the \textit{active ingredients} thought to promote change, the specific \textit{targets} to be changed, and the hypothesized \textit{mechanism(s)} by which the active ingredients influence the targets. The inclusion of the hypothesized mechanism(s) means that therapies documented according to the RTSS will automatically suggest causal mechanisms to incorporate into our causal model (see~Section). The RTSS also encourages documenting treatment targets more precisely than common practice; for example, rather than treating ``gait,'' a target might be to improve step length symmetry or regularity. This complements the Novel~Measurement~Approaches we described, which make routinely measuring changes in these targets feasible.

Another important point emphasized by the RTSS is the connection to Enablement Theory \parencite{whyte_contributions_2014}, which links changes between the domains of the ICF. In the RTSS, therapy goals are referred to as \textit{aims}, which may reflect performance measured in the Activity and Participation domains of the ICF downstream of the treatment target. Similarly, our framework seeks to computationally operationalize Enablement Theory by tracking changes in function across the domains of the ICF, quantifying their interdependencies, and building causal models of their changes over time.

The RTSS also addresses the volitional component of many interventions as a causal modulator that is critical for their real-world effectiveness, with some active ingredients designed to increase volitional behavior \parencite{whyte_importance_2019}. Many approaches to telerehabilitation automatically allow capturing real-world performance, thus providing insight into the effectiveness of volitional ingredients that can be incorporated into our causal models. We elaborated on this more in our case study on gamified therapy (see EMG~Biofeedback~for~Arm~Recovery~After~Spinal~Cord~Injury).

RTSS only provides a high-level framework, so delineating many of these components to document concretely is a substantial endeavor currently underway. Additionally, the burden of documenting all active ingredients and dosing parameters is in tension with the practicalities of providing efficient clinical care. Furthermore, perhaps even the exact kinematics of individual trials are important to capture. We anticipate that advances in computer vision and activity recognition will enable this level of precise documentation of interventions to be automatically generated from the observation of treatment sessions, although this will require the development of dedicated algorithms to capture these features.

\subsection{Computational Neurorehabilitation}

Our framework is inspired by and builds upon the work of \textcite{reinkensmeyer_computational_2016} on Computational Neurorehabilitation. Like the RTSS, Computational Neurorehabilitation emphasizes the hypothesized mechanisms of interventions and proposes building quantitative longitudinal models. Ideally, these should be neuro-mechanistic models of the activity-dependent plasticity that underlies recovery after a neurologic injury. They note the challenges in modeling multiple parallel phenomena, including 1) capturing the process of spontaneous recovery that still may be modulated by activity; 2) modeling the influence of the activity-based therapies on recovery at the trial-to-trial level on recovery, ideally at the granularity of individual trials; and 3) decomposing functional recovery into restitution, reflecting repairing the original neural circuits or functions, and compensation through learning new strategies. They emphasize the opportunities for collecting the data to build these models from robotic therapies and wearable sensors (see Novel~Measurement~Approaches).

\textcite{reinkensmeyer_computational_2016} focuses on the recovery of the upper extremity after stroke and points out that the existing models are still overly simplistic, such as modeling the change in strength at a single joint. Eight years later, this limitation still generally persists. We hope the additional components of our framework that facilitate the pragmatic application to large heterogeneous datasets will reduce this limitation. Additionally, our framework more explicitly focuses on bridging functional levels across the ICF to link changes in impairments to changes in what matters to patients -- function in the real world -- and does this through structured causal models and causal inference.


\subsection{Novel study designs}

\textcite{french_precision_2022} indicated the importance of synergistic study designs to advance precision rehabilitation. A strength of the causal framework we are proposing is that it can learn from a mixture of heterogeneous datasets ranging from longitudinal observational data to randomized clinical trials. In the current state of rehabilitation, there is often clinical equipoise over a range of possible rehabilitation interventions, resulting in practice variation between and within sites that the causal models can learn from. However, this data may still be insufficient to answer all causal questions. CI provides tools to determine when causal effects require additional interventional data for identification or when there are wide confidence bounds on intervention effectiveness. In these cases, additional interventional studies may be required to identify the likely causal model. Furthermore, a challenge for rehabilitation research is how long most interventions must be administered to produce measurable and clinically meaningful differences. Many rehabilitation interventions are more effective as combinatorial treatments, such as adding neuromodulation to a high-dose therapy with additional motivation components. Adaptive trial designs may be particularly synergistic with our framework to pragmatically provide the answers needed for rehabilitation practice and precision rehabilitation.

The Multiphase Optimization Strategy (MOST) is an efficient and structured framework for optimizing interventions and disentangling the effect from combinatorial interventions prior to a definitive clinical trial \parencite{collins_optimization_2018}. This optimization stage is often neglected in rehabilitation research, particularly when considering the constraints on real-world clinical rehabilitation. Adaptive study designs can accelerate this process. For example, Microrandomization Trials (MRT) can be used to systematically and rapidly vary certain active ingredients in interventions while measuring engagement, dosage, or changes in movement kinematics \parencite{klasnja_microrandomized_2015}. Sequential Multiple Assignment Randomized Trials (SMART) \parencite{almirall_introduction_2014, collins_optimization_2014} allow the introduction of multiple points of randomization and criteria for crossing over to test interacting hypotheses and optimize interventions. The RTSS notes that progressing the intensity of therapies is often necessary, keeping them easy enough to avoid failure but difficult enough to promote recovery \parencite{hart_theory-driven_2019}. Testing the specific adaptive progression rules is a case where SMART trials can provide concrete evidence and help identify the ODTR \parencite{lei_smart_2012, almirall_introduction_2014}.

\subsection{Causal Inference}

Causal inference (CI) is an analysis framework that can analyze the big rehabilitation data with the components described above \parencite{kaddour_causal_2022}. An accessible introduction to CI we recommend is the ``Book of Why'' \parencite{pearl_book_2018}, and interested readers can refer to the Appendix for a brief primer on some pertinent concepts. We will very briefly describe CI and how it relates to our framework.

The distinction between CI and traditional statistics is that in CI, analyses include both data and a hypothesized causal structure of the data-generating mechanisms that produce the data. This can be formalized as a diagram indicating the causal interactions between variables, such as using Structural Equation Models. These causal models then make available new analytic methods, termed do-calculus \parencite{pearl_causality_2009}, that can estimate the causal effects between variables. This reflects the type of inferences we ultimately want, answering whether specific interventions would change specific outcomes for a particular patient. CI tools can also determine whether there is sufficient data to identify this causal effect or whether additional, possibly interventional, data would be needed. Thus, causal analyses can provide much deeper insights from data than analyzing without formalizing the causal hypotheses.

CI complements the goal of \textcite{reinkensmeyer_computational_2016} to use Computational Neurorehabilitation to model the plasticity underlying recovery. In our framework, we propose to build more comprehensive models that capture recovery of impairment and functioning across levels of the ICF and over time. \textit{Building these causal models is not easy}. There will also not be just one causal model. Rather, they will need to be created for specific conditions and questions. This will require a close interplay between domain experts, clinical stakeholders, and data scientists. CI can also be applied to cross-sectional data to identify interactions. For example, an elegant study by \textcite{schwartz_model_2022} on gait laboratory data from nearly 10k pediatric participants used CI to model the interaction between diagnoses, specific impairments of body structure and function and their impact on gait mechanics, mobility, energy expenditure, and quality of life. In section Unpacking~the~Causal~Box, we provide two example causal models we are developing for longitudinal studies. Furthermore, there are exciting developments in using data to identify the causal variables and their relationships, termed causal representation learning and causal discovery, which we believe will be essential to untangle the complex interdependencies of the rehabilitation process.

\section{Causal Framework for Precision Rehabilitation}

\begin{figure}[!htbp]
\centering
\includegraphics[width=1\linewidth]{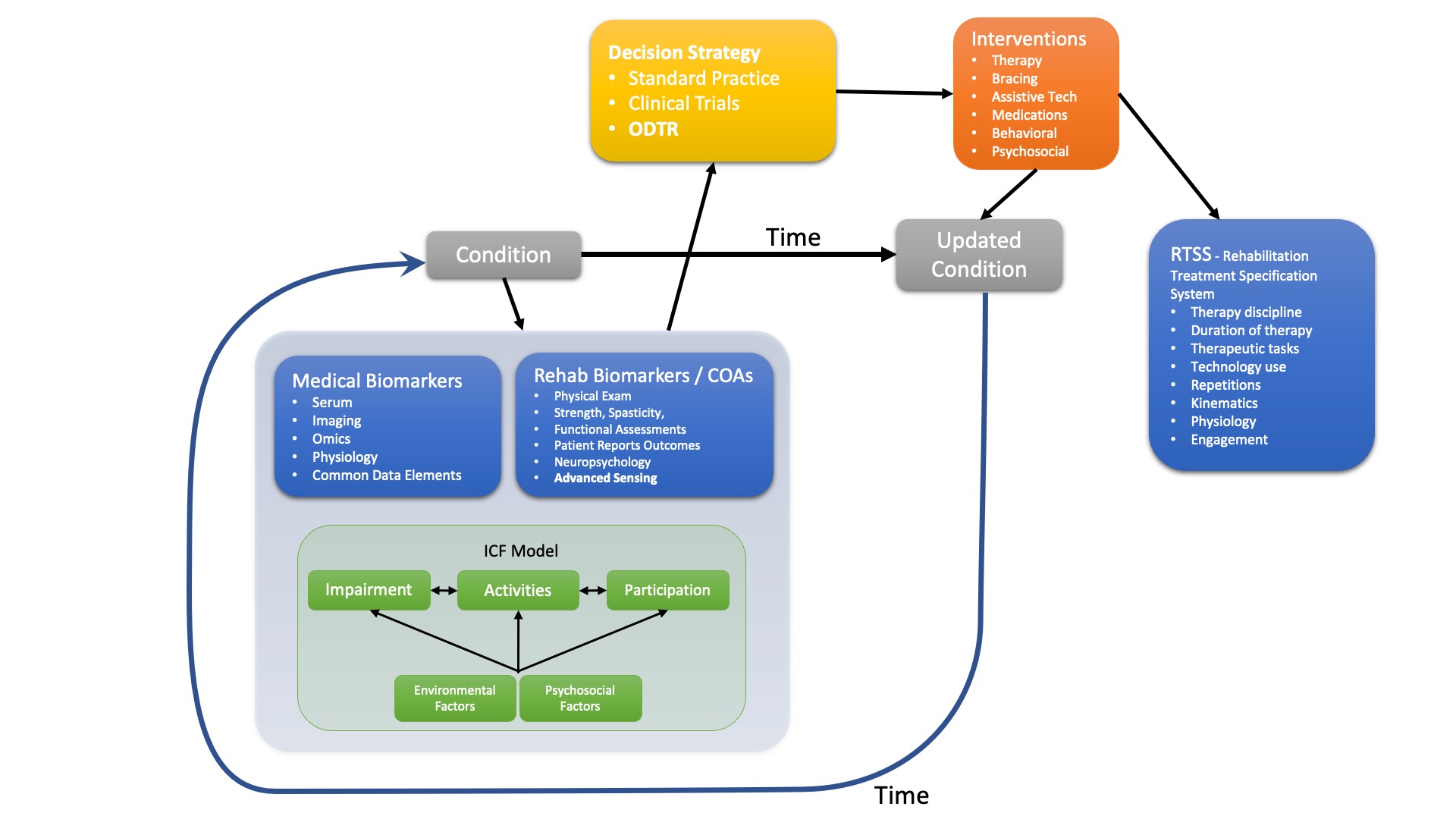}
\caption[]{High-level diagram describing the process that generates rehabilitation data. At any given time, the patient has an underlying condition, reflecting the state of their body and support system. This gives rise to a range of measurements we can make, including biomarkers and clinical outcome assessments (COAs) that can be measured to gain insight into this, which are contextualized with the ICF. Then, interventions are selected based on the measurements and decision strategies, ranging from standard practice to clinical trials or even the ODTR. Ideally, these are precisely documented per the RTSS. As a result of these interventions, the patient has an updated health condition, and the process repeats.}
\label{fig:data_generation}
\end{figure}

Here, we describe how the previously enumerated components can be integrated into our Causal Framework for Precision Rehabilitation. Figure~\ref{fig:data_generation} provides a high-level diagram of the rehabilitation process that generates the data that will power precision rehabilitation. We will collect longitudinal datasets describing how patients' health conditions evolve. This includes medical biomarkers, including lab tests, physiologic mapping, omics, and imaging, although we do not discuss these at length in this work. The datasets must also include social determinants of health, which greatly influence rehabilitation outcomes. It also includes rehabilitation biomarkers and COAs, which can be augmented by movement measurements from computer vision and wearable sensors. The datasets should contain detailed characterization of interventions, which will be documented using the RTSS to describe the active ingredients and treatment targets. Much of the work to productively use our framework requires developing more causal models appropriate for the question of interest (see Unpacking~the~Causal~Box). In this diagram, health condition is represented as a single box but reflects a collection of latent variables describing the state of the neuromusculoskeletal system and the psychosocial context that influences function. We will briefly linger at this level of abstraction to highlight several important points.

First, the biomarkers and COAs we measure are causally downstream of the latent variables underlying the patient's health condition. However, they rarely provide direct measurements of those variables. This is analogous to the ICF delineating different domains, with COAs designed to measure function within those constructs, albeit sometimes imperfectly. In some cases, measurements may more closely correspond to specific variables in the body (e.g., neuroimaging), but in other cases may have a more complex relationship and be influenced by multiple internal variables (e.g., step width variability). Formalizing the link between causal latent variables and biomarkers (or COAs) offers a principled way to integrate information from multiple partially redundant biomarkers and identify the most informative biomarkers. This also will facilitate a deeper understanding of the causal relationships between underlying constructs and the outcome of interest.

Second, a key contrast between precision rehabilitation and precision medicine is the need to consider the patient through the lens of the ICF model. To enable precision rehabilitation, the model must capture the longitudinal influence caused by our interventions and how they relate across these levels. Some interventions may improve participation via improving repair at the neural level, which reduces impairment and thus improves functioning \parencite{krakauer_broken_2017}. Other interventions may use assistive devices and training to improve compensation. Compensation can benefit function at the activities and participation level without reducing body structure and function impairments. Social factors are prominent in the ICF and have a complex and substantial causal influence on outcomes. Modeling these will require patient-provided data about other potential causal influences outside the formal healthcare systems. It is critical for precision rehabilitation that the research and analysis is precise about these mechanistic pathways to understand how they interact to optimize the prescription of interventions.

An accurate and quantitative understanding of the links between levels of the ICF will improve the sensitivity and specificity of our research studies and guide the use of novel biomarkers. Rehabilitation clinicians, patients, and their advocates ultimately want improvements at the activity and participation levels, but producing these improvements often takes a long time, and many measures have limited sensitivity, construct validity, or accuracy. In contrast, changes in impairment can typically be measured more precisely, such as strength measured with dynamometry or muscle activation with electromyography. Changes at the impairment level might be detectable sooner than changes in function, given these changes in impairments often precede and mechanistically cause improvements at the activity and participation level. Thus, more sensitive impairment measures can better power and accelerate research. However, those impairment changes may not matter to patients if they don't improve valued dimensions of real-world function. A robust understanding of how much changes in impairment influence functioning will guide the use of measures of impairment as surrogate biomarkers \parencite{fda-nih_biomarker_working_group_best_2016} for functional recovery in clinical trials and monitoring responses to interventions.

After building causal models of the appropriate granularity for particular clinical and research questions, which we elaborate on in the next section, we will use CI to fit them to the data described in Figure~\ref{fig:data_generation}. These models will serve as Digital Twins \parencite{bjornsson_digital_2019, kamel_boulos_digital_2021, masison_modular_2021} for \textit{in silico} experiments to identify the ODTR Figure~\ref{fig:odtr}. Because the causal models predict function across the ICF and over time, the ODTR can also be trained to maximize the attainment of the patient's goals on a longer time scale. This modeling can predict if a combination of interventions will produce changes in outcomes that are meaningful to the patient. Conversely, they may identify that some interventions reduce impairments but not to a level that produces clinically meaningful results in activities or participation, which should thus be further improved through research before clinical dissemination.

\subsection{Unpacking the Causal Box}\label{causal_box}

The key challenge of this framework will be to unpack the ``Condition'' box of Figure~\ref{fig:data_generation} into a causal structure that can be modeled and fit with the available theory and supportive data. This causal structure can computationally formalize rehabilitation questions into a unifying framework that can bridge the functional levels of the ICF and span the time course of recovery. Ideally, the causal models should reflect specific mechanistic models of plasticity. They should include modeling changes that occur in the absence of formal interventions (sometimes called spontaneous recovery) and could even capture the causal influence of each therapeutic repetition driving a small amount of activity-dependent plasticity in the nervous system, while scaling up to how those plastic changes accumulate to improve function. However, in most cases, we are far from having such mechanistic models, particularly ones link from the intricacies of the nervous system to participation in the community. We anticipate a hierarchy of models from more abstract latent variable representations learned through work with domain experts. Advanced analysis tools can further augment these models using techniques like causal representation learning and causal discovery (see Details~on~Causal~Inference).

There are several known constraints. As constructed, measures of impairment in body structure and function typically quantify specific components of the neuromusculoskeletal system. Many therapies are designed to explicitly treat impairments, such as increasing strength, reducing spasticity, and increasing range of movement, many of which we can quantify. Hypothetically, reducing impairment should causally improve function in the activities and participation domains. However, in many cases, there are redundant ways to accomplish tasks. Alternatively, a patient may learn new ways to accomplish tasks to improve performance in activities and participation domains without changes in impairment, often termed compensation. With improvements in kinematic monitoring of patients both in the clinic and the community, these distinct outcomes can be characterized and tracked. Other rehabilitation interventions improve function despite stable impairment, such as providing an ankle-foot-orthosis to improve ankle and knee control.

These models can then be extended with more detailed neuro-mechanistic components. For example, if the primary mechanism of an intervention is to strengthen corticospinal connectivity, measurements such as motor-evoked potentials can validate this mechanistic hypothesis while mathematically modeling expected changes in a functional movement. Recent advances in neural and neuromuscular modeling in animals show the promises of this approach for studying plasticity \parencite{dewolf_neuro-musculoskeletal_2024}, or for incorporating models of the spinal reflex circuits \parencite{bhattasali_neural_2024}. Because of the transportability of causal models, a properly designed model can be fit to a mixture of data, including subsets that include these more detailed physiological measurements, to learn a more robust model that generalizes across subjects. They can even provide a coherent model that can be fit to mixtures of human and non-human data.

To make this framework more concrete, we present two case studies we have been developing within this precision rehabilitation framework. The first is modeling arm recovery after spinal cord injury using biofeedback games, and the second is an application to gait rehabilitation after stroke. For clarity, we use feedback loops to capture plastic changes through time rather than building a strictly structured causal model. Both examples are under development and have yet to be fully realized. We also expect any particular diagram will likely frustrate some domain experts when it neglects their favorite hypotheses, and we use this as an opportunity to emphasize that the diagram serves to make such hypotheses explicit and testable via CI.

\subsubsection{EMG Biofeedback for Arm Recovery After Spinal Cord Injury}\label{gamified_biofeedback}

Gamified EMG biofeedback provided via wearable sensors is a therapy with a pathway to broad dissemination that could help people with spinal cord injury independently achieve high dosages of therapy and ultimately improve arm function. However, it is a long way from a single trial of muscle activation and the limited plastic change it might induce to clinically meaningful improvements in arm function. EMG biofeedback likely works through two primary mechanisms; in the short term, it can help overcome components of learned disuse by showing participants the activation of muscles when they otherwise cannot detect it. At the longer time scale, it likely works by promoting activity-dependent plasticity that strengthens residual connections below the level of injury. The premise is that a high enough dosage drives enough plasticity to produce improvements in arm strength that result in improved function in the real world.

\begin{figure}[!htbp]
\centering
\includegraphics[width=1\linewidth]{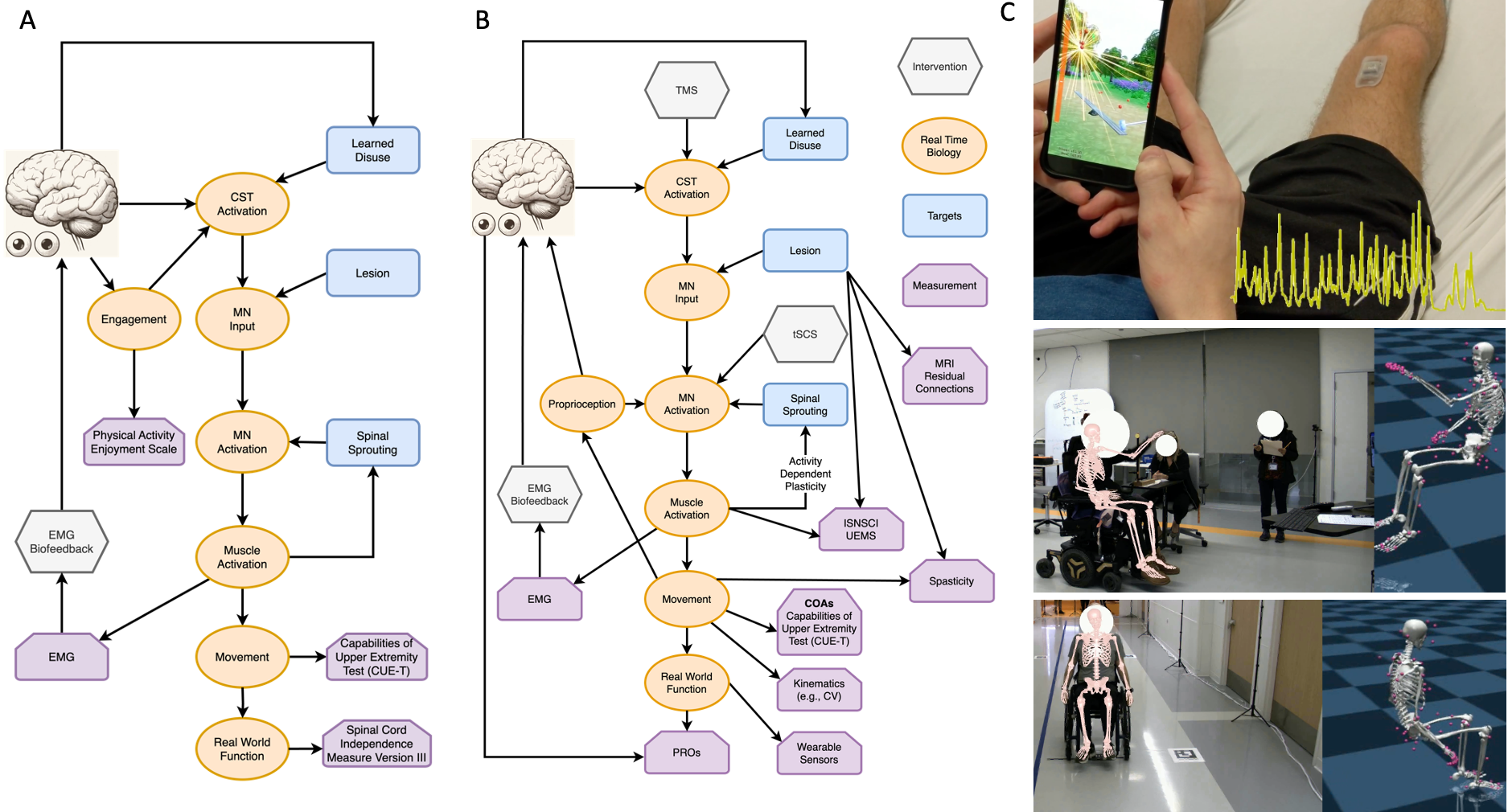}
\caption[]{Causal diagram for EMG biofeedback program. A) shows a more minimal model that includes some of the targets of this therapy, including learned disuse and spinal sprouting. It captures how EMG is measured and provided as biofeedback to induce muscle activation that can alter these targets. It also captures how these latent variables influence other measurements, such as arm function and real-world independence. B) shows a more comprehensive model that includes additional measurements, such as kinematics or measures that reflect spinal cord injury lesions. C) includes images of some of these measurements, including the EMG measured during biofeedback or arm kinematics during upper extremity assessments and wheelchair propulsion.}
\label{fig:emg_bfb}
\end{figure}

Figure~\ref{fig:emg_bfb} unpacks a hypothesized causal structure underlying this therapy and with the latent variables in the nervous system that may drive the observable measurements. The minimal model (left) shows how EMG biofeedback promotes muscle activation to alter spinal sprouting treatment targets and learned disuse. Muscle activation also produces the movement scored during the Capabilities of Upper Extremity Test (CUE-T) \parencite{marino_development_2012} and the arm function in the real world, assessed with the Spinal Cord Independence Measure Version III (SCIM-III) \parencite{ackerman_using_2010} and Patient Reported Outcomes (PROs).

The full model (middle) adds strength assessments from the International Standards for Neurological Classification of Spinal Cord Injury (ISNCSCI). It also adds arm movement kinematics during assessments, wheelchair propulsion, or other activities at home (e.g., \parencite{bandini_measuring_2022}). It can be extended to include the interaction of transcutaneous spinal cord stimulation to improve the recruitable motor unit pool for biofeedback, which would be immediately apparent from the EMG. The model can capture neurophysiological tools like transcutaneous magnetic stimulation to measure residual corticospinal connectivity. Alternatively, corticospinal connectivity can be inferred from more clinically accessible tools such as spasticity \parencite{sangari_spasticity_2023} or MRI assessments of residual fiber bridges in the cord \parencite{freund_mri_2019}. Additional components could be naturally added to this model. For example, there is the opportunity to adjust the EMG biofeedback difficulty per trial based on the measurements, and this can be tuned to balance difficulty and therapeutic efficacy with engagement and duration of therapy. Alternatively, this diagram could be further extended to include the coordination between muscles in the generation movement and could add an intervention for coordination using kinematic biofeedback from wearable sensors.

After fitting this model to sufficient data, the goal would be to learn the ODTR that improves real-world function based on PROs such as the SCIM-III. This model could also determine what level and severity of lesion indicates that EMG BFB will likely result in a clinically meaningful benefit to the participant or if the therapy should be further augmented, such as with tSCS.

\subsubsection{Gait Rehabilitation After Stroke}\label{gait_rehabilitation}

\begin{figure}[!htbp]
\centering
\includegraphics[width=1\linewidth]{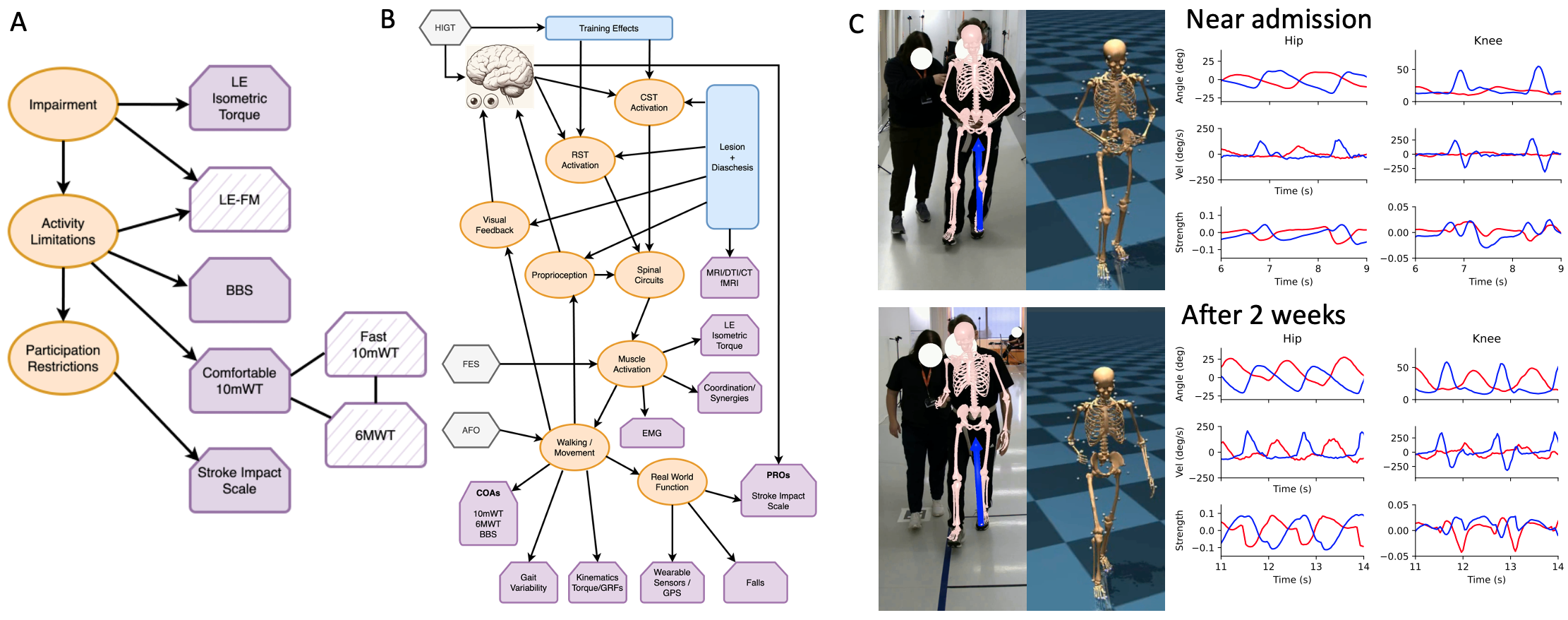}
\caption[]{Example of causal framework applied to post-stroke gait impairments. A) results from \textcite{sullivan_model_2011} with the causal interactions modeled more abstractly at the level of the ICF. Hashed elements were those excluded from the final model. B) is a more neuroanatomically motivated model that captures the multiple pathways and feedback loops responsible for gait impairments. This model also indicates how interventions like high-intensity gait training (HIGT), functional electrical stimulation (FES), and ankle-foot orthoses (AFOs) can be incorporated. C) shows an example of how computer vision can extract a great deal of information about changes in walking during inpatient rehabilitation for someone with a stroke, with our latest methods even inferring ground reaction forces and net joint torques. This shows increases in the independent modulation of the paretic limb joint torques.}
\label{fig:stroke_gait}
\end{figure}

The second case study we present is post-stroke gait recovery. This is challenging as mapping from factors in the nervous system to the complex regulation of observable walking is complex, and no comprehensive models describe this process. \textcite{sullivan_model_2011} explored a causal perspective for post-stroke gait but focused on the causal interactions between the constructs of impairment, activity restriction, and participation limitations. They found impairment influenced participation only indirectly, via a direct influence on activity, a similar result to work in knee osteoarthritis \parencite{pollard_exploring_2011}. They then performed confirmatory factor analysis to verify that COAs and other measurements captured their intended construct. Their results are reflected in Figure~\ref{fig:stroke_gait}A. The factory analysis \textcite{sullivan_model_2011} revealed that the Berg Balance Scale (BBS) loaded onto activity instead of impairment, as originally predicted, and the Fugl-Meyer Lower Extremity score loaded equally onto impairment and activity instead of impairment alone. This speaks to the challenges of the psychometrics of existing COAs to cleanly partition into the ICF domains. They also found the fast 10-meter walk test (10mWT) and the 6-minute walk test (6MWT) to be too colinear with comfortable 10mWT to be included as independent measurements. However, from the perspective of our framework, this approach is lacking. Critically, ``impairment'' is not a causal variable in of itself, but a grouping of observations arising from causal variables in the nervous system.

Figure~\ref{fig:stroke_gait}B shows a potential, more neuro-mechanistic model that accounts for how the lesion and training effects alter the relative control of walking from the corticospinal tract (CTS) and bilateral reticulospinal tract (RST). This is significant, as it is believed that greater reliance on the RST leads to more dominant synergy patterns, with some hypothesizing that excessive early task-specific training may strengthen the RST at the expense of the overall long-term recovery potential of the CST \parencite{krakauer_broken_2017}. Synergy patterns can be detected through clinical assessments, EMG measurements, or even through kinetics during walking from computer vision. Furthermore, the node for muscle activation can be further extended to include spinal reflexes using neuromuscular biomechanical modeling \parencite{caggiano_myosuite_2022}. For example, \textcite{song_evaluation_2017} uses such a reflex model for walking, albeit with a highly simplified model of the brain. Similarly, models with changes in spinal reflex circuitry can reproduce gait patterns of other diseases, such as Hereditary Spastic Paraplegia \parencite{lassmann_dysfunctional_2023}. This model also neglects many brain regions that control gait and can be impacted by a stroke, which are reviewed in \textcite{beyaert_gait_2015}, and ultimately, these regions should be included and linked to neuroimaging data describing the lesions.

The model in Figure~\ref{fig:stroke_gait}B is still vague about how therapies influence recovery of the CST or dependence on the RST and would need to be extended to capture spontaneous recovery and processes like diaschesis, as well as neuromuscular modeling to capture processes like spasticity. However, such a model would be of significance to patients and clinicians if it could show that a therapy that requires an early overdependence on RST pathways not only influences joint synergy patterns but also negatively impacts long-term gait speed, gait endurance, falls, and ultimately participation. Such trials are already underway for upper extremity recovery after stroke.

\subsection{Optimal Dynamic Treatment Regimen}

What we ultimately want for precision rehabilitation is not only a causal model (e.g., Figure~\ref{fig:emg_bfb} or Figure~\ref{fig:stroke_gait}) that provides a good fit to our data (e.g., Figure~\ref{fig:data_generation}), but also the ODTR (e.g., Figure~\ref{fig:odtr}). This formalism of precision rehabilitation provides a data-driven suggestion for the intervention that will maximize the patient's function, participation, and goal achievement over their lifespan \parencite{hurn_goal_2006, choi_goals_2017}. The ODTR can be obtained from the learned causal model by using it as a Digital Twin. Digital Twins have been adopted from engineering to precision medicine \parencite{bjornsson_digital_2019, kamel_boulos_digital_2021, masison_modular_2021}, and provide a digital model of a patient that enables high-throughput, \textit{in silico} experiments to optimize interventions. The causal models we propose will produce such Digital Twins for rehabilitation against which we can optimize interventions to maximize long-term function. There are multiple ways this could be accomplished.

One approach could use CI to infer the value of the unobservable variables in the causal diagram from the available measurements. Then, the search space of possible interventions could be exhaustively searched with rollouts over time to find a strategy that maximizes the patient's function. The optimal next interventions could then be applied. After obtaining additional measurements, this process would be iteratively repeated. This approach of searching for the optimal next action, applying it, and repeating the search using updated measurements has close links to optimal control methods, like Model Predictive Control.

In some cases, the search process for optimal control may be too computationally expensive to run for each patient at each point in time.  Another approach for learning decision-making policies is Reinforcement Learning (RL) \parencite{sutton_reinforcement_2018}. Causal Reinforcement Learning \parencite{zhang_designing_2020} applies RL to causal models and can learn a decision-making policy equivalent to the optimal dynamic treatment regimens. By appropriately specifying the reward signal, the trained ODTR will optimize an outcome under the experimenter's control. This means the ODTR can designed to maximize long-term function rather than reduce short-term impairments.

In this discussion of our framework, we focused on using causal reinforcement learning against a structured causal model fit to longitudinal data, as this provides a cohesive framework that can embrace mechanistic computational neurorehabilitation. It is worth briefly mentioning this is not the only approach to learning an ODTR, and other CI tools like the potential outcomes (PO) framework attempt to directly learn the ODTR from data \parencite{tsiatis_dynamic_2019}. When applied to iterative decision-making, these methods must causally account for the confounding influence of prior decisions on current observations. For example, causal trees that do this can learn iterative decision rules for ICU ventilator management \parencite{blumlein_learning_2022}. Another approach that may be easier than fully structured causal models is generative latent models, which have recently shown success in modeling the complex longitudinal trajectories for systemic sclerosis \parencite{trottet_modeling_2023}. Like the causal models we have described, these can predict future outcomes conditioned on interventions and historical data while not requiring all causal mechanisms to be formally stated.

Finally, once an ODTR is identified, the causal models indicate whether it will improve outcomes compared to the standard of care. If so, the model can be deployed in clinical trials to validate its effectiveness while keeping humans in the loop. This is similar to the intervention optimization and testing workflow recommended in MOST \parencite{collins_optimization_2018}. These clinical trials will remain essential. While one goal of causal modeling is to improve generalization and resilience to changes in the data distribution, deploying the system may change enough things that some of the model assumptions are violated. Such clinical trials naturally keep clinicians in the loop, with the ODTR simply suggesting interventions predicted to be optimal. Testing the ODTR will also produce data that further refines the fitted causal model performance and lead to iterative improvement in the learned ODTR.

\subsection{Data-driven Identification of Biomarkers (Phenotyping)}

A prerequisite for an ODTR to be effective is having enough measurements to phenotype a patient and determine the optimal intervention. This is increasingly complex with the burgeoning set of measurements one can obtain from medical biomarkers, rehabilitation biomarkers, COAs, and AI-powered measurements with emerging technologies. Furthermore, many measurements do not provide independent information, such as the comfortable 10mWT, fast 10mWT, and 6MWT in \textcite{sullivan_model_2011} all being colinear and loading onto the activity factor (Gait~Rehabilitation~After~Stroke). Given logistical constraints, obtaining all possible measurements at each point in time to guide therapy is infeasible. Rather, we want the most informative measurements that produce actionable information that results in better outcomes.

Fortunately, the causal models and tools from CI can transform this question into one that is computationally tractable and answerable within our framework. Ultimately, we want to know how much an additional measurement reduces our uncertainty on the value of the latent variables in the causal diagram and how much that will influence the optimal interventions via the ODTR. Thus, we envision our framework will provide a more data-informed way to guide the collection of laboratory data, imaging, and performing COAs. This may result in algorithms for performing assessments analogous to computer-adaptive testing, where the next most informative questions are iteratively determined, producing more accurate assessments in less time. Algorithms to recommend assessments can also be constrained to consider cost, time, and what is feasible for a patient to achieve safely. For example, when assessing balance, in some cases, performing a tandem walk with eyes closed under cognitive loading may be required to expose performance limitations, but for some people, that would immediately result in a fall. The most informative and safe next test depends on the current estimate and uncertainty of stability. Similarly, in spinal cord injury research, multiple approaches can be used to assess residual corticospinal connectivity, including motor-evoked potentials, advanced neuroimaging, and even quantitative spasticity assessments (EMG~Biofeedback~for~Arm~Recovery~After~Spinal~Cord~Injury). These all relate to similar underlying latent variables. Thus, the causal models provide a way to integrate these assessments and, within the broader framework, quantitatively compare their contributions as different classes of biomarkers. The causal modeling framework can also help to identify the classes of biomarkers that measurements serve as, such as diagnostic, response, prognostic, or predictive biomarkers, or whether certain measurements can serve as surrogate endpoints \parencite{fda-nih_biomarker_working_group_best_2016}.

In parallel, we anticipate using technologies like wearables and computer vision to produce significantly more information from any given COA. For example, our algorithms and workflows allow us to begin integrating 3D gait analysis into the 10-meter walk test (Figure~\ref{fig:stroke_gait}C), one of the most commonly performed COAs for mobility. This is motivated by our belief that a careful analysis of movement patterns combined with big data will provide substantial and valuable insights into the nervous system's functioning and recovery process. Our framework provides a cohesive way to integrate these new sources of information, test this belief, and keep the focus on how to use this information to enable better outcomes.

\section{Discussion}

\subsection{Comparison to French et al. (2022)}

Much of what we propose aligns closely with the \textcite{french_precision_2022} vision of precision rehabilitation and builds upon it. To make this explicit, we endorse the value of the critical components they enumerated:

\begin{enumerate}
\item Synergistic use of various study designs, including RCTs and observational studies. We also discussed the potential value of quality improvement projects and adaptive study designs. By comparing model fits and looking at the confidence intervals over causal effects, our framework can guide studies to where model uncertainty remains high but could meaningfully improve quality of life.
\item The need for standardized functional measurements. This is crucial to better compare across datasets. However, by predicting the performance of functional measurements from latent causal variables, our framework is more agnostic to the set of functional measurements. It can also identify the optimal set of measurements, both to improve model fitting and to deploy in clinical practice for phenotyping and subtyping patients. We also advocate for the development of common metrics through the use of standardized and reproducible units of measurement.
\item The importance of precise and longitudinal measures of function. We fully agree, including on the use of computer vision and wearable sensors. We add that understanding the construct validity of these measures will give us equations by which to test underlying mechanisms for treatment targets. We also emphasize the importance of detailed measurement of interventions to parallel this data, ideally documented according to the RTSS with the treatment targets, treatment ingredients, and hypothesized mechanisms of actions \parencite{hart_theory-driven_2019, van_stan_rehabilitation_2019}.
\item The utility of comprehensive datasets. A well-curated dataset with meaningful performance metrics to the machine learning community often stimulates rapid advances. However, a barrier to even anonymized datasets is the risk of re-identification (see \parencite{chikwetu_does_2023} for cases with wearable sensors). Federated Learning and Synthetic data may help learn from multiple datasets while lessening these barriers \parencite{li_federated_2020, nguyen_novel_2022, hesse_learning_2018}.
\item The importance of predictive analyses. Our framework fully embraces this idea, focusing on mechanistically understanding how changes occur across different constructs within the ICF and on the need to predict longitudinal function and participation. This can lead to the ODTR, which suggests interventions to help people achieve their individualized long-term goals, thus ensuring this predictive analysis has value to end users.
\item The need for system and team science. We agree with this and, in particular, emphasize the need to collaborate with researchers proficient in causal inference to fit our models and with rehabilitation researchers to identify hypothesized causal mechanisms to incorporate into these models.
\end{enumerate}

\subsection{Comparison to Reinkensmeyer et al. (2016)}

Similarly, our work strongly builds upon \textcite{reinkensmeyer_computational_2016} proposal to build models of recovery. We extend their perspective to modeling across all levels of the ICF. Our framework also places a greater emphasis on using data to identify novel or combinatorial intervention strategies to improve patient-centered outcomes while highlighting impairment as an important level of measure and treatment in service of these outcomes. We suggest methods to analyze large heterogeneous clinical datasets with the tools of CI to leverage big data to understand the precision influence of interventions. We also emphasize the role of the RTSS in providing documentation sufficient to achieve this. Finally, we identify the ODTR as both the formalisms of the key tool for precision rehabilitation and describe how it can be derived from causal models.

\subsection{Future Directions and Challenges}

It should be apparent that this framework cannot immediately be applied to easily available public datasets and used to improve outcomes. It will require first mobilizing the relevant data from the silos of proprietary electronic medical records and ensuring this documentation contains the necessary information. It will further require diligent work to develop causal models at the appropriate granularity to use this data best to produce clinically actionable or testable insights. As noted, CI with appropriate models can benefit from the heterogeneity of datasets between clinical sites, but leveraging this data will be most feasible using federated learning approaches, which will require collaboration on common infrastructure and frameworks to implement practically.

These challenges will require the rehabilitation research community to establish strong connections with the data science and causal inference communities. This will allow the rehabilitation community to communicate the complex nature of rehabilitation to data scientists and formulate the problems that matter to our patients into computationally tractable ones. It will also allow the data scientists to inform the rehabilitation teams where certain questions cannot be answered with the available data and to determine where we need more explicit interventional studies or to add more measurements into rehabilitation practice.

To make this framework more concrete, we largely focused on motor rehabilitation after neurologic injury, and for these examples, we only discussed a subset of the causal factors that must be considered to fully model recovery. The framework can be naturally extended to other conditions and a more comprehensive consideration of the patient. For example, the ICF highlights the interaction of psychosocial factors with impairment, activities, and participation. Any practicing rehabilitation clinician is acutely aware that these causal factors dominate long-term outcomes through many pathways. For example, having reliable transport to therapies, having the appropriate home modifications and DME to allow them to leave their home, and whether there is a team in place to prevent secondary complications, just to name a few examples. Thus, it will be essential to model these factors. Additionally, psychosocial factors interact with mental health, which itself has a myriad of pathways to influence recovery. Some of these might be observable and measurable, such as engagement with therapies, but others might require additional psychological assessments. Management of depression and anxiety is common during rehabilitation, and tracking these interventions can also be incorporated into the causal models. These all further interact with cognitive impairments, changing how people engage and the efficacy of specific therapeutic interventions.

Furthermore, this framework can be extended to maintaining functional ability with aging without active issues,  musculoskeletal conditions, or developmental conditions.  In the case of developmental and some neurologic conditions, we anticipate the role of genomics and other molecular biomarkers to be particularly important and can be incorporated into the causal models.

Our proposed framework is well-suited to maintaining and improving functional ability across life stages, including aging and the onset of other medical and psychosocial challenges.

\section{Conclusion}

Despite the daunting challenges, we are optimistic that a precision rehabilitation framework can materially improve rehabilitation strategies and clinically meaningful outcomes. We are at an inflection point for AI-driven advances to allow us to collect much richer data throughout the rehabilitation process and the real world that has been sorely lacking. However, rehabilitation currently lacks a systematic framework to transform this data into improved clinical decision rules. A multidisciplinary and patient-inclusive team science application of causal inference to these data offers the opportunity to obtain greater mechanistic insights into recovery and can identify the Optimal Dynamic Treatment Regimens for individual patients. This framework can help rehabilitation teams maximize the obtainment of patient-valued long-term goals.

\section{Acknowledgments}

The Medical Rehabilitation Research Resource (MR3) Network National Coordinating Center is funded in part by a grant from the NIH/NICHD/NCMRR -- award number P2CHD086844. RJC is supported by the Craig H. Neilsen Foundation, the Restore Center P2C (NIH P2CHD101913),  R01HD114776 and the Research Accelerator Program of the Shirley Ryan AbilityLab.

\printbibliography

\if@endfloat\clearpage\processdelayedfloats\clearpage\fi

\begin{appendix}
\begin{appendixbox}
\subsubsection{Details on Causal Inference}\label{appendix}

Causal models are commonly represented as a diagram with arrows indicating which variables influence other variables, which is a directed acyclic graph (DAG) (e.g., Figure~\ref{fig:emg_bfb} and Figure~\ref{fig:stroke_gait}). This is accompanied by equations that describe how each variable is influenced by variables with arrows going to it and by sources of noise, called the Structured Causal Model (SCM). A nuance of this structure is that measurements, such as gait kinematics or rehabilitation outcome measures, are \textit{outputs} of the causal model, which are influenced by internal latent variables. These latent variables may have straightforward neurophysiologic interpretations, such as strength in a muscle or preservation of corticospinal connectivity, with the former being an example of a variable that is also much easier to measure directly. However, in other cases, the variables may be more constructs aligned with the ICF, such as ``mobility impairment''. Thus, causal models can serve as a tool to investigate the construct validity of different outcome measures (For an example, please see Gait~Rehabilitation~After~Stroke). CI includes methods to infer the values of latent variables in the causal diagram from the observed measurements \parencite{zheng_using_2019}, such as do-calculus.

The do-calculus \parencite{pearl_causality_2009} behind CI uses the causal diagram to compute the strength of the causal effects from these datasets. Appropriately specified models naturally control for confounding factors (within the scope of the proposed mechanisms), such as disentangling who provided interventions from what the intervention was. Measuring confidence intervals on causal effect explicitly tests hypotheses such as ``given this structure and data, is there evidence that intervention X has a benefit for the average participant or even for a particular type of participant.'' This subquestion is particularly important as it distinguishes between the average treatment effect (ATE) and conditional average treatment effect (CATE). The ATE is the average effect of an intervention across all participants, while the CATE is the average effect for a particular subgroup. Identifying the biomarkers and delineating patient subgroups with a greater CATE for specific interventions is the essence of precision rehabilitation.

A significant challenge to the application of CI to rehabilitation is that we are in the infancy of understanding the complex interactions of all the pertinent causal elements underlying recovery. In the ideal case for CI, one knows the causal diagram and can directly measure all of the variables -- the opposite of our case. Tools to make CI applicable to large data with missing measurements and imperfect understanding of the causal structure are improving. This includes approaches like Causal Representation Learning, which tries to learn the causally relevant variables, and Causal Discovery, which learns the causal graph between these variables (e.g., \parencite{wendong_causal_2023}). This approach can be seen as a generalization of Independent Component Analysis (ICA), as it models the causal dependence among latent components, focusing solely on learning the unmixing function and the causal mechanisms. \textcite{locatello_challenging_2018} challenges common assumptions in the unsupervised learning of disentangled representations, indicating the fundamental impossibility of this task without inductive biases on both models and data. Moreover, as \textcite{fumero_leveraging_2023} suggests, leveraging sparse and shared feature activations can enhance disentangled representation learning, particularly when applied to real-world data, which has traditionally been overlooked in favor of simpler synthetic settings.

Once the pieces are in place - big longitudinal multisite datasets combined with novel measurements and interventions documented by the RTSS - we envision a program of transdisciplinary collaboration that incorporates the known features of computational neurorehabilitation and also uses causal representation learning and discovery to identify clinically meaningful causal factors underlying recovery. These learned factors can then be tested and validated. As this process continues, it will advance from abstract latent variable representations to concrete neurobiologically and socially grounded models.

There are several CI concepts worth highlighting for how they map onto a framework for precision rehabilitation. The first concept is called the Ladder of Causality or Pearl's Causal Hierarchy \parencite{pearl_book_2018}. This corresponds to a hierarchy of questions that can be answered within a CI framework. The first level contains association questions, corresponding to a traditional approach of measuring correlations. The second level includes interventional questions, such as those addressed through RCTs, which answer ``do outcomes improve if given this treatment''. The third level addresses counterfactual questions, which imagine if the causal structure were modified what would happen to answer ``why?'' questions about what might have happened if a different treatment plan was administered to an individual. Answering these second and third-level questions from large rehabilitation datasets would power precision rehabilitation to improve outcomes.

Another CI concept is identifiability. Identifiability is a process for answering ``given this causal diagram and this data, can I support or refute my hypothesis?'' This goes beyond a power analysis, and addresses, given the causal structure and available data, does the set of interventions and measurements provides sufficient insight into the system to address this question. Identifiability can determine whether additional measurements or structured changes in interventions are necessary to answer the precision rehabilitation hypothesis. This can guide decision-making for trial and analysis designs, such as what additional variables must be extracted from the electronic medical record or what aspects of practice need to be documented in greater detail to test their efficacy.

Another concept is transportability, which means that a causal model can be applied to different datasets that may have systematic differences, such as different levels of impairments or medical complexity, because the underlying mechanisms are the same. Transportability also allows hypotheses to be tested by aggregating data between sites, making certain hypotheses identifiable that would not be from either dataset alone. For example, if there are different practice patterns between two clinical sites. This process has close links to propensity matching and other approaches that can address confounders when working with large data. Propensity matching was used to answer the causal question of whether discharging patients with a stroke from acute care to either skilled nursing facilities or inpatient rehabilitation facilities improves outcomes. The analysis of data from more than 100,000 people supported the hypothesis that inpatient rehabilitation results in improved functional outcomes \parencite{hong_comparison_2019}. Through causal methods, this study answered an important practice question that could not be answered with an RCT.

A significant challenge to applying Causal Inference (CI) to rehabilitation is that we are only beginning to understand the complex interactions of all the pertinent causal elements that underlie recovery. Ideally, in CI, one would have a clear causal diagram and the ability to directly measure all of the variables involved. However, our situation is quite the opposite. Tools to make CI applicable to large data with missing measurements and an imperfect understanding of the causal structure are improving, including approaches like Causal Representation Learning, which aims to infer causally related latent variables along with the unknown graph encoding their causal relationships \parencite{wendong_causal_2023}. This approach can be seen as a generalization of Independent Component Analysis (ICA), as it models the causal dependence among latent components, focusing solely on learning the unmixing function and the causal mechanisms. Another promising avenue is Causal Discovery, which focuses on identifying the causal graph between variables. In this regard, the work by Locatello et al. challenges common assumptions in the unsupervised learning of disentangled representations, indicating the fundamental impossibility of this task without inductive biases on both models and data \parencite{locatello_challenging_2018}. Moreover, as Fumero et al. suggest, leveraging sparse and shared feature activations can enhance disentangled representation learning, particularly when applied to real-world data, which has traditionally been overlooked in favor of simpler synthetic settings \parencite{fumero_leveraging_2023}. Once these pieces - big longitudinal multisite datasets combined with novel measurements and interventions documented by the Rehabilitation Treatment Specification System (RTSS) - are in place, we envision a program of transdisciplinary collaboration that incorporates the known features of computational neurorehabilitation and also utilizes causal representation learning and discovery to identify clinically meaningful causal factors underlying recovery. These learned factors can then be tested and validated. As this process continues, it will advance from abstract latent variable representations to concrete neurobiologically and socially grounded models.

\subsubsection{Is the model good enough?}

There is essentially an unlimited space of models that can be fit to longitudinal rehabilitation data, so how can we them down? Model comparison provides a quantitative approach to compare how well different causal models fit the data, allowing us to test causal hypotheses and determine which model better explains the longitudinal data. It also provides an alternative to RCTs, which are both expensive and rarely provide a cohesive account for all of the data collected.

Models can also be compared using heterogeneous datasets combined from different sites with different standard-of-care treatment patterns and even different standardized measurements. These can be further augmented with datasets from RCTs with intentionally randomized interventions. Furthermore, CI algorithms can determine whether causal hypotheses are even testable with the existing data or require additional interventional studies, allowing this computational framework to help design research studies.

After fitting a model to the data, we can make two useful types of predictions. The first is to obtain confidence intervals on the strength of specific causal interactions, enabling the previously described hypothesis testing for the presence of causal interactions and to estimate the effect size. More importantly, the models predict the functional benefits of different counterfactual treatment programs. This can distinguish therapies that might produce a measurable change in impairment but that will not meaningfully change a person's real-world function from those that will. In the former case, these may be promising research interventions in need of further optimization or combination with other strategies.

Ultimately, after the hard work of building and fitting these models is performed, additional hard work will be required to validate them. For example, \textcite{schwartz_model_2022} addresses this in their causal model of gait impairments by comparing many of the identified causal interactions to known biomechanical principles, providing a validation of the face validity of the inferences. Validation of the predictions from the ODTR will ultimately need prospective clinical trials that are carefully monitored with humans in the loop.
\end{appendixbox}
\end{appendix}


\end{document}